\documentclass[a4paper,
              ]{paper}

\usepackage{graphicx}

\begin{document}

\title{Effect of Beam Dynamics Processes in the Low Energy Ring ThomX\thanks{This work is supported by the French "Agence Nationale de la Recherche" as part of the program "investing in the future" under reference ANR-10-EQPX-51. This work was also supported by grants from R\'egion Ile-de-France.}}

\author{ N. Delerue\thanks{delerue@lal.in2p3.fr}, C. Bruni, I Chaikovska, I. Drebot\thanks{illya.drebot@mi.infn.it now at INFN-Milan, Italy},\\  M. Jacquet, A. Variola, F. Zomer\\ {\it Laboratoire de l'Acc\'el\'erateur Lin\'eaire (LAL),  Universit\'e Paris-Sud XI,}\\ {\it F-91898  Orsay, France}\\ \\
	    A. Loulergue, {\it Synchrotron SOLEIL, St Aubin, France}}

\maketitle

\begin{abstract}
As part of the R\&D for the 50 MeV ThomX Compton source project, we have studied the effect of several beam dynamics processes on the evolution of the beam in the ring. The processes studied include among others Compton scattering, intrabeam scattering, coherent synchrotron radiation. We have performed extensive simulations of a full injection/extraction cycle (400000 turns). We show how each of these processes degrades the flux of photons produced and how a feedback system contributes to recovering most of the flux.
\end{abstract}

\section{Beam dynamics at ThomX}
ThomX~\cite{ThomX2011} will be a 50~MeV compact accelerator used to produce X-rays by Compton scattering between electrons and a laser. It will be made of a linac and a ring in which the beam will be stored for 20~ms. A more detailed description of the project is available elsewhere in these proceedings~\cite{ThomX2014} and in the projet's TDR~\cite{ThomX_TDR}. Given the very short storage time, the short bending radius and the low energy of the electrons, several beam dynamics effect will have a strong effect on the expected flux.

An extensive study of the beam dynamics at ThomX has been performed, including among others the effects of Compton backscattering (CBS), intrabeam scattering (IBS) and coherent synchrotron radiation (CSR). This study is described in~\cite{ThesisIllya} and we present here its main results.
 
\section{Simulation code}

\begin{figure}[!htb]
    \centering
    \includegraphics[width=70mm]{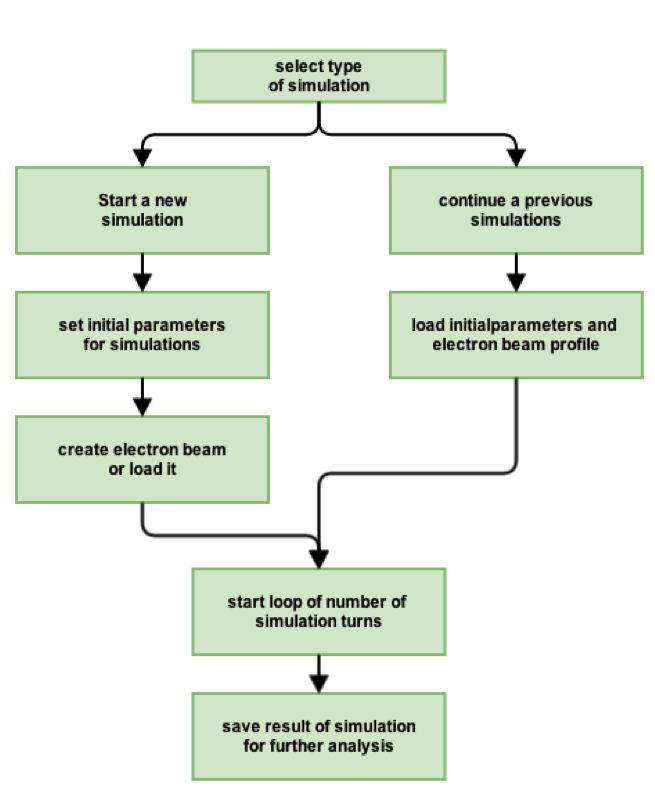}
    \includegraphics[width=70mm]{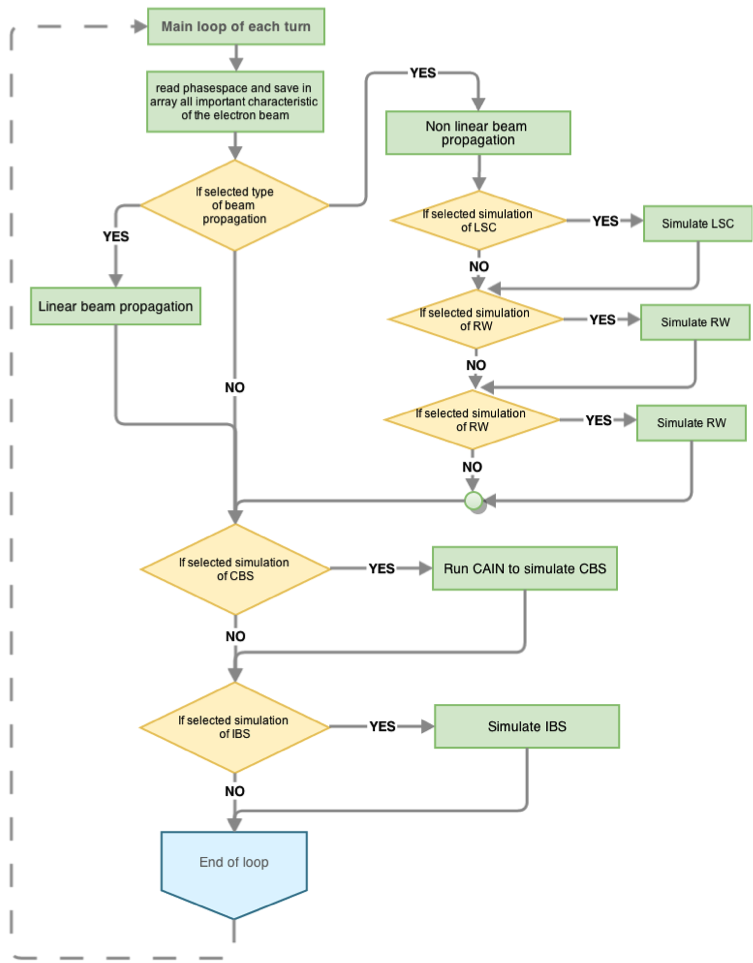}

    \caption{Structure of the simulation code. Top: overall structure. Bottom: Simulation loop.}
    \label{structure_code}
\end{figure}

The simulation code we have used is based on  Matlab. It is written as a shell in which several beam dynamics effects are implemented with macro-particles beings passed from the simulation of one effect to the other as arrays. The code has also the flexibility of calling external program such as CAIN~\cite{cain}. The properties of the macro-particles can be exported after each turn although it is advised to do it less often to minimize the disk space used. The structure of the code is shown on figure~\ref{structure_code}. 

\begin{figure*}[!!!tbh]
    \centering
    \hspace*{-10mm}\begin{tabular}{ccc}
    \includegraphics*[width=50mm]{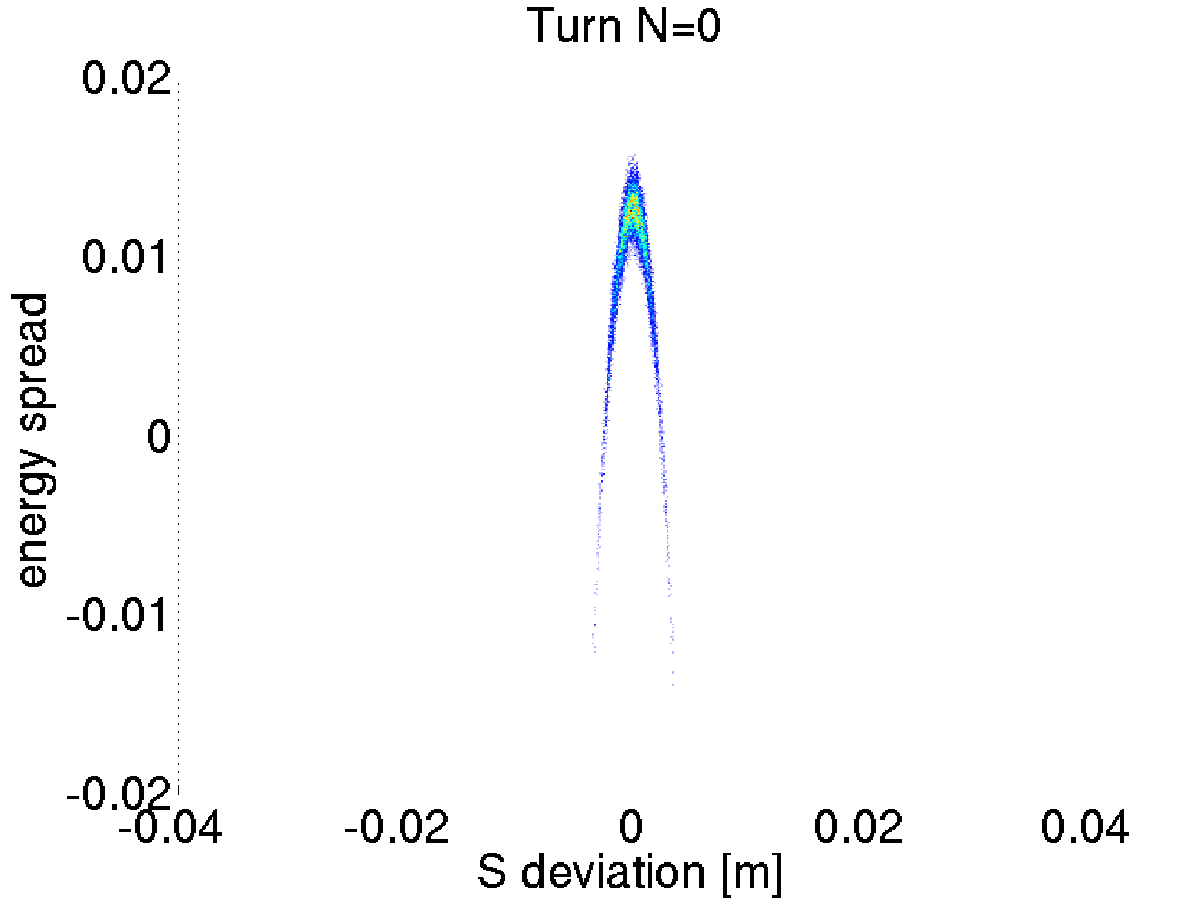}  &
    \includegraphics*[width=50mm]{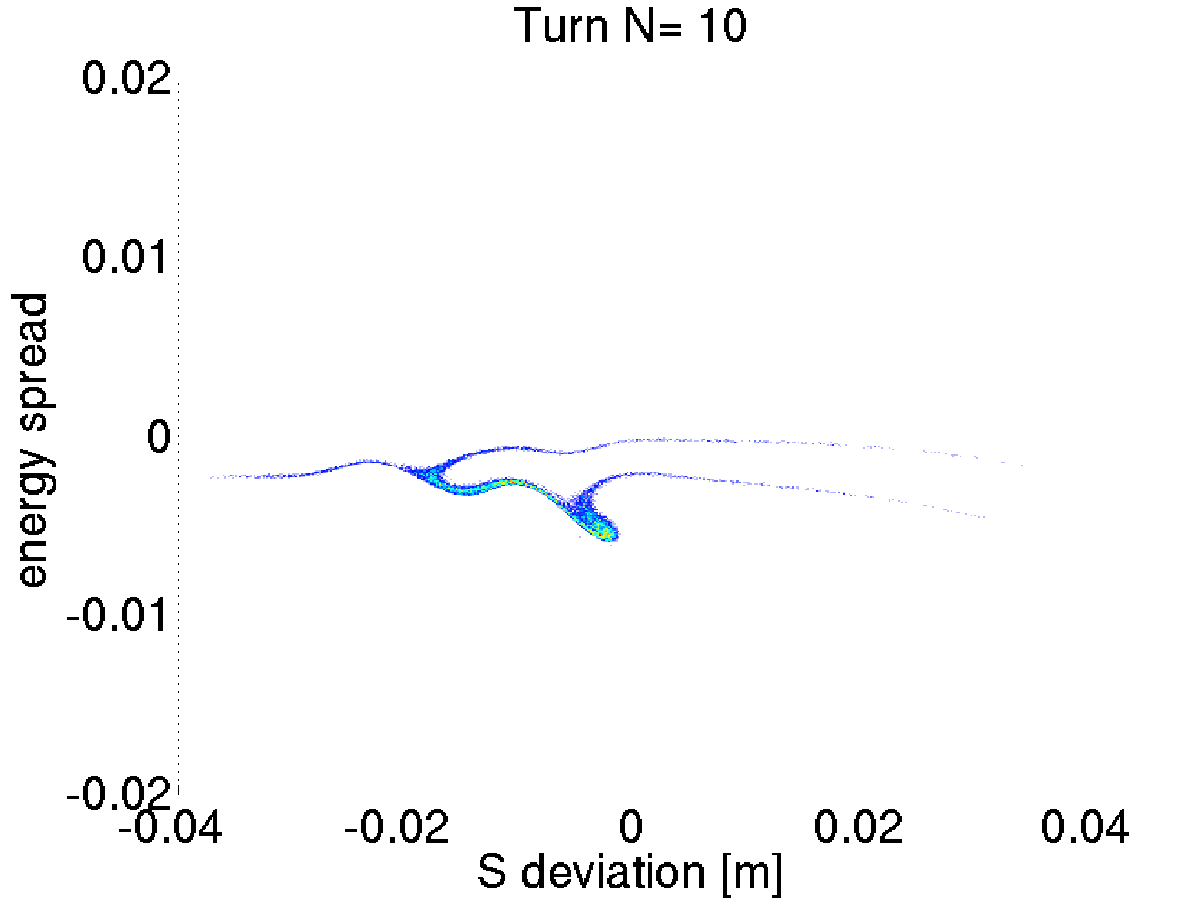}  &
    \includegraphics*[width=50mm]{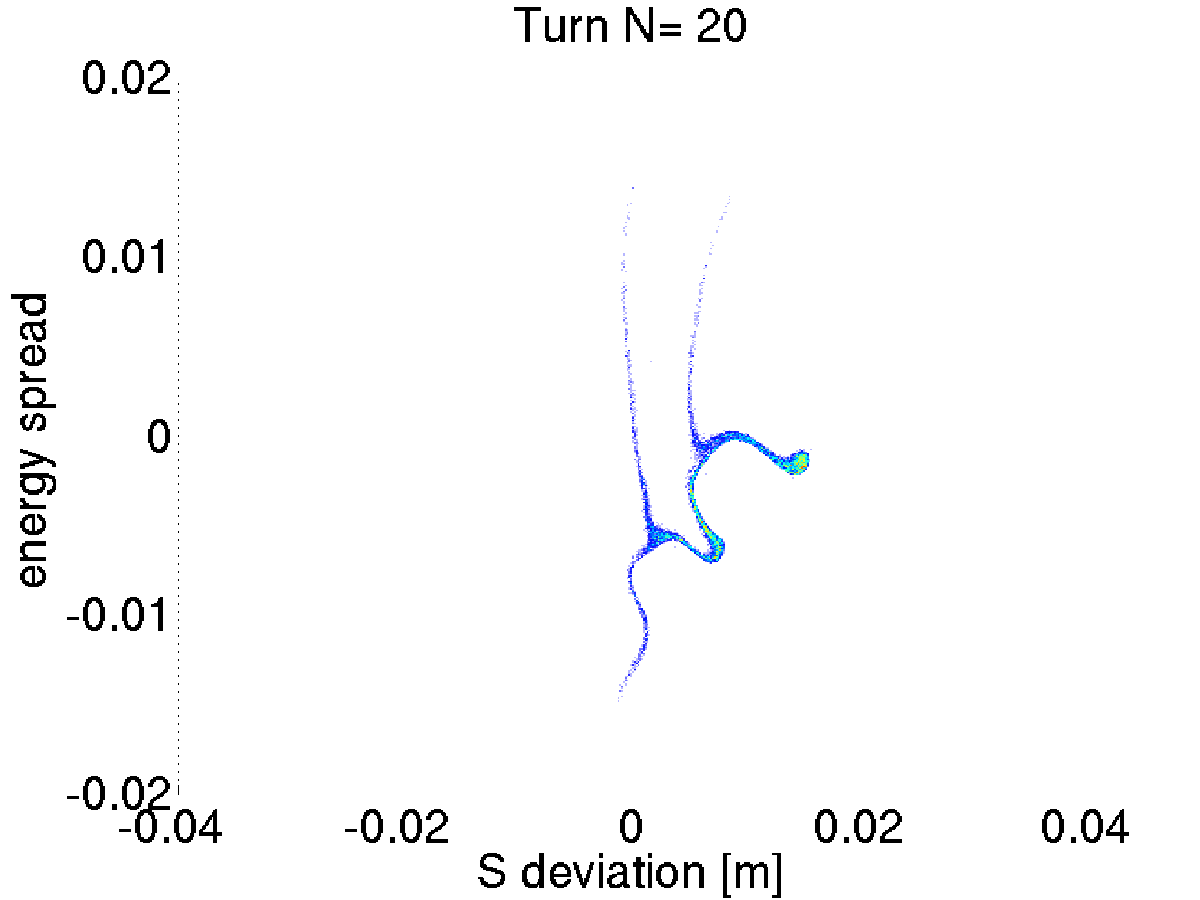} \\
    \includegraphics*[width=50mm]{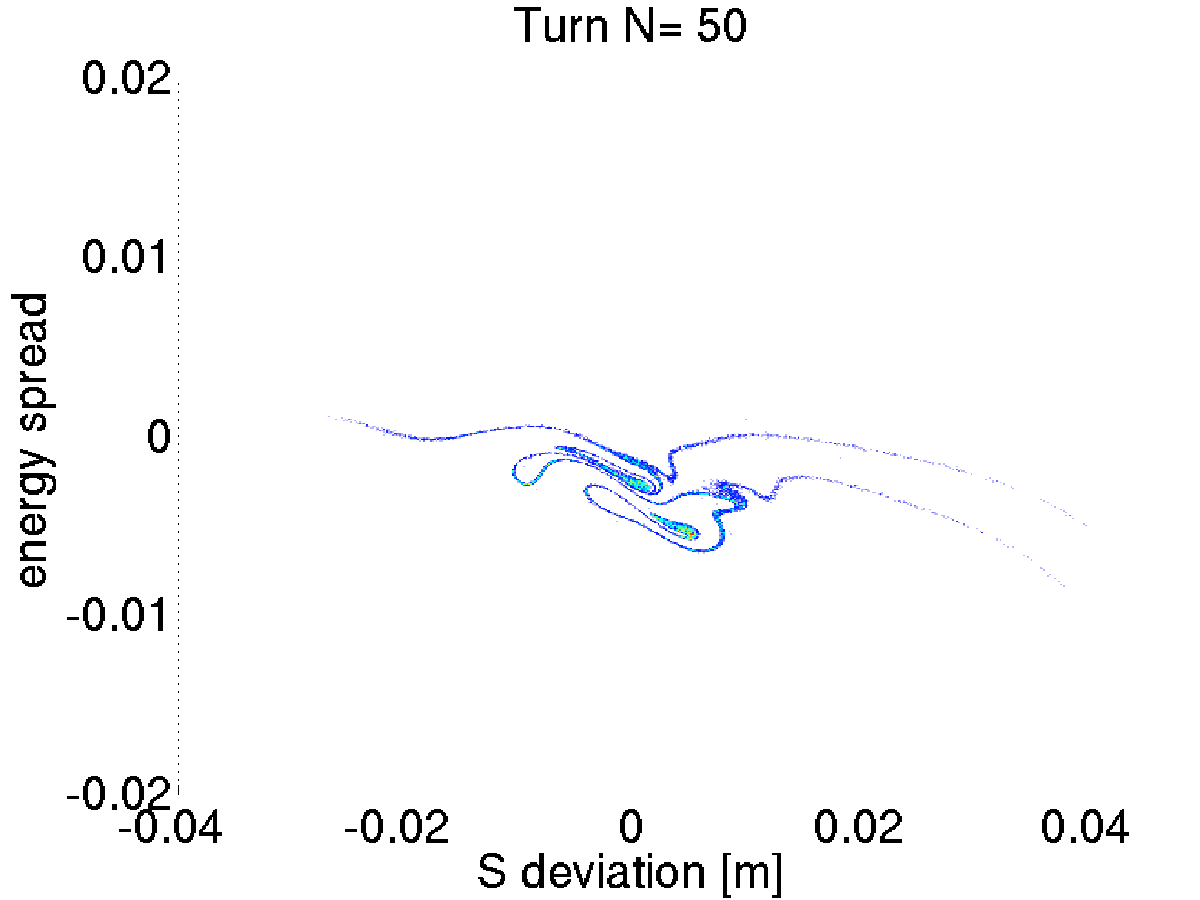}  &
    \includegraphics*[width=50mm]{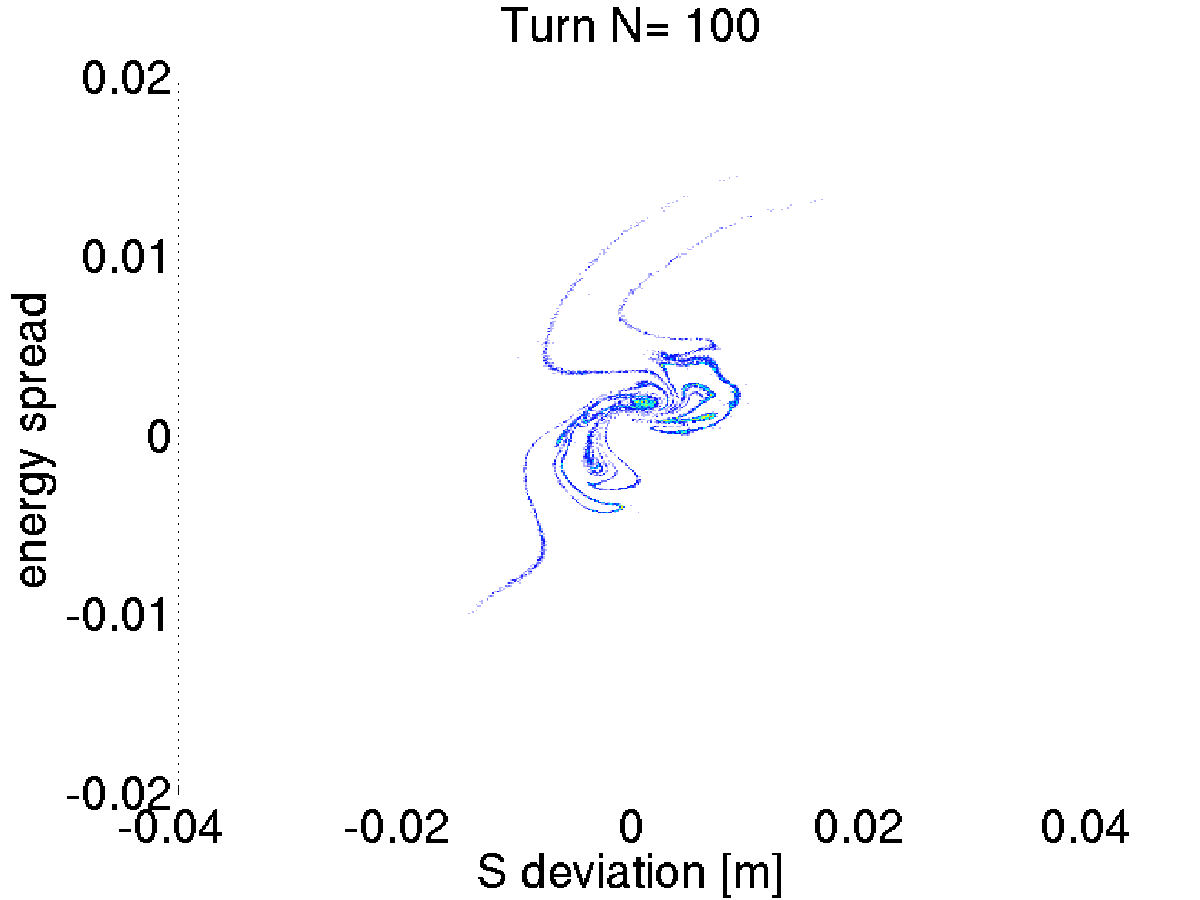} &
    \includegraphics*[width=50mm]{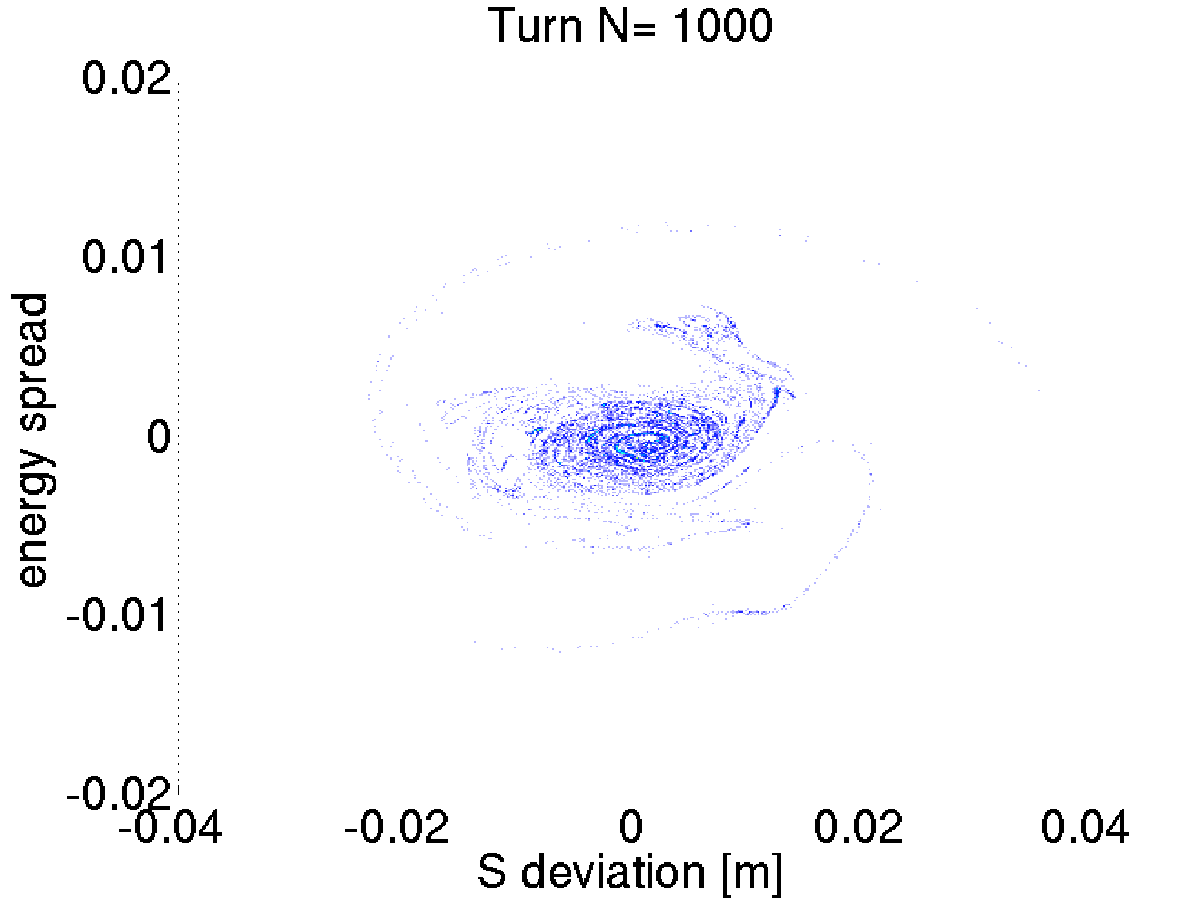} 
   \end{tabular}
    \caption{Evolution of the bunch in the longitudinal phase space (energy spread versus longitudinal position) as function of turn number (indicate above each image).  The color of the dots indicate the particles intensity at that location red being more intense than green and green more intense than blue. }
    \label{evolution}
\end{figure*}

The possibility of exporting all macro-particles properties also allows to break the execution of the simulation in several jobs to be executed serially on a computer farm. The duration of the simulation will depend on the number of macro-particles chosen. For $10^5$ macro-particle a simulation of the full injection/extraction cycle will take several week although, as we will see below, this can be significantly shortened if one is mostly interested in what happen during the initial transient period.

\section{Initial instabilities due to mismatching}

Using the code with have made a full simulation of the evolution of the longitudinal phase space as function of time. As expected and as shown on figure~\ref{evolution} just after injection the beam enters in a turbulent regime. This is due to the shape of the injected beam being mismatched with the shaped of the bucket. During this turbulent regime a fraction of the charge may be lost.

\section{Effect of Coherent Synchrotron radiation}

\begin{figure}[!htb]
    \centering
    \includegraphics[width=90mm]{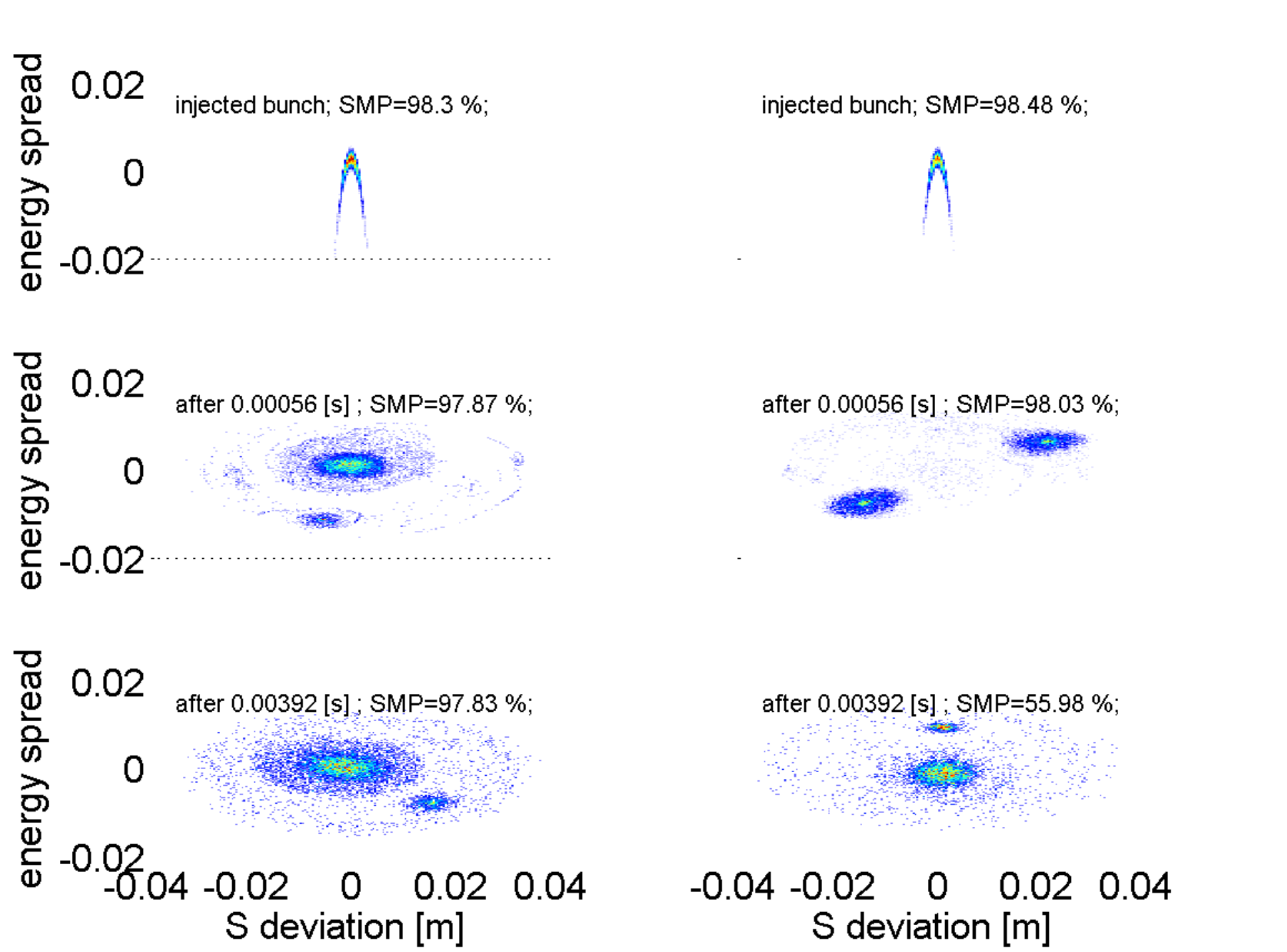}

    \caption{Longitudinal phase space of the bunch (energy spread versus longitudinal position), at injection (top), after 10~000~turns (middle) and after 70~000~turns (bottom) in two cases with the same beam parameters (but different initial random distributions of the macro particles). On the left  the bunch is relatively stable and only 2\% of the macro particles are lost whereas on the right the bunch is unstable and it is split in less than 10~000~turns, leading to the loss of 44\% of the charge. The color of the dots indicate the particles intensity at that location red being more intense than green and green more intense than blue.}
    \label{fig:CSR_split}
\end{figure}

At high charge the most disruptive effect is CSR. While performing our simulations we saw that sometimes CSR can split the bunch in two or more parts leading to a significant fraction of the charge to be ejected from the bunch. This result in a lower X-ray yield. Examples of simulations of the longitudinal phase space with and without
such beam break-up are shown on figure~\ref{fig:CSR_split}: for the same initial beam parameters but for different random distributions of the macro particles one case leads to a relatively stable beam whereas the other case leads to the bunch been split in two parts within less than 10~000~turns and one of the two parts is then lost shortly after leave only 56\% of the macro particules in the bunch.

\begin{figure}[!htb]
    \centering
    \includegraphics*[width=90mm]{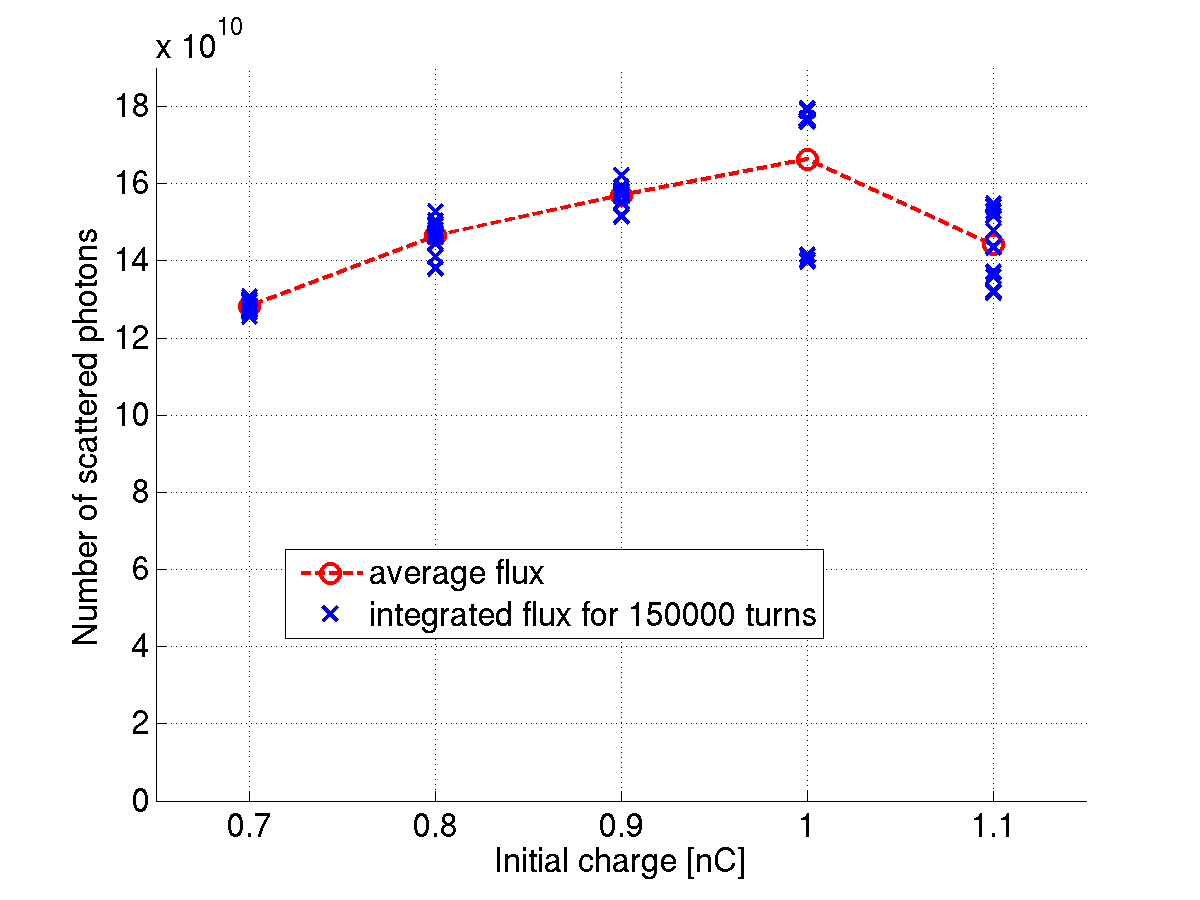}

    \caption{Average photon flux produced for 10 simulations of 1500~000 turns each for 5 different initial charges. At 1nC and above during some simulations the beam get partially lost leading to a lower photon yield.}
    \label{sum-flux}
\end{figure}

This beam break-up means that there is a threshold above which adding charge at injection will in fact decrease the charge that can actually be stored and therefore decrease the X-ray yield. This is shown on figure~\ref{sum-flux} where one can see that the average flux produced for an initial charge of 1nC is higher than the flux produced for an initial charge of 1.1nC.

\section{Overall flux}

\begin{figure}[!htb]
    \centering
    \includegraphics*[width=90mm]{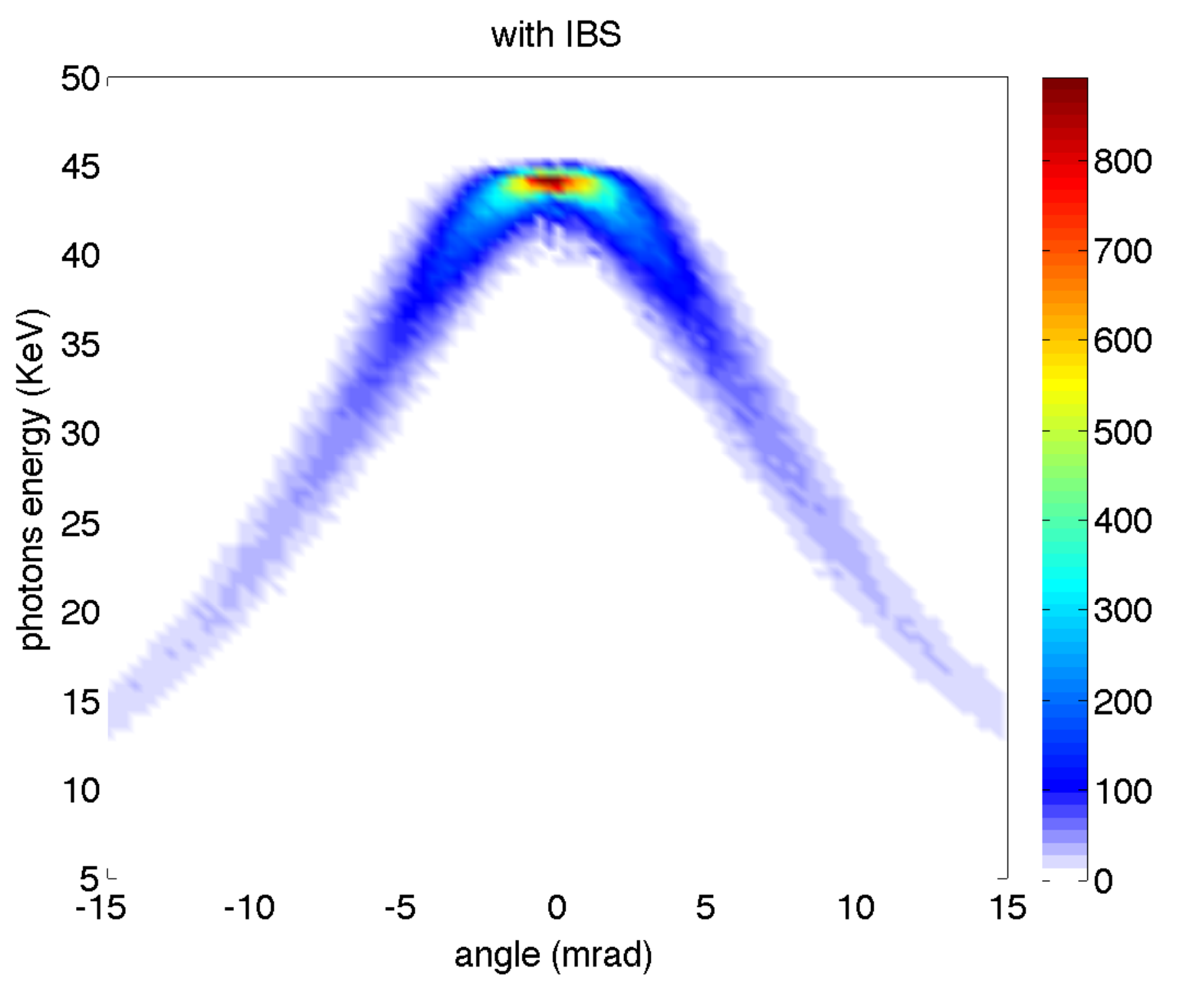}

    \caption{Distribution of the scattered photons as function of their energy and scattering angle. The color indicate the photon intensity at a given coordinate point from blue (less intense) to red (more intense). }
    \label{spectrum}
\end{figure}

Using the simulation code it is possible to simulate a full storage and predict the spectrum obtained. Fig.~\ref{spectrum} shows an example of photons spectrum where the intensity is plotted against the energy (vertical axis) and the scattering angle (horizontal axis). 

\begin{figure}[!tbh]
    \centering
    \includegraphics*[width=90mm]{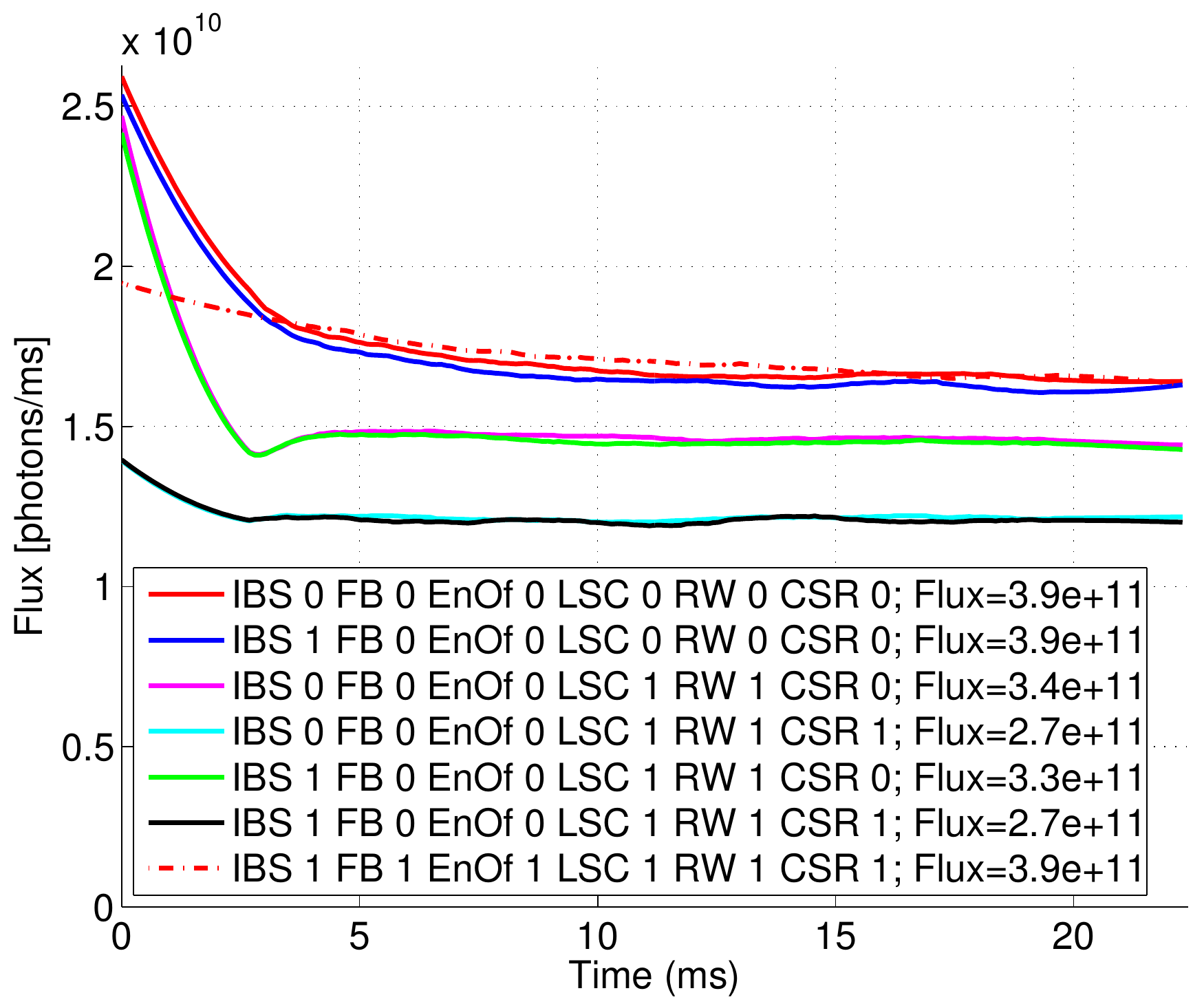}

    \caption{Photon flux as function of the turn number with various effects turned on or off. In the caption, IBS stands for Intrabeam Scattering, FB  for longitudinal feedback, EnOf for Energy Offset (to compensate beam mismatch), LSC for longitudinal space charge, RW for resistive wall effect, CSR for Coherent Synchrotron Radiation. When an effect is followed by a 0 it means that it is turned off. If it is followed by a one, it means that it is turned on in the simulations. The flux in the caption is given in number of photons  (integrated over the duration of the simulation). }
    \label{compare_effects}
\end{figure}

It is also possible to investigate the effect of each beam dynamics effect on the final flux. For that we made several simulations where we turned on or off some effect selectively to see how doing this changed the flux. The result of these simulations is shown on figure~\ref{compare_effects}. Although in reality it will not be possible to  turn off effects, understanding what is causing the losses will help us in designing the best possible mitigation strategy.

\section{Outlook}

We have written a simulation code to study the beam dynamics at the ThomX Compact Light Source. This allows us to better understand how beam dynamics will affect the photon flux with the aim of designing a mitigation strategy. We see that CSR is the effect that will the most significantly decrease the flux and the main reduction in flux will occur during the first 5~ms of the cycle.

%


\end{document}